\documentstyle[prl,aps,twocolumn,epsfig]{revtex}

\begin{document}

\preprint{}
\draft

\title {The Edge of Quantum Chaos}

\author{Yaakov S. Weinstein$^{\dagger}$, Seth Lloyd$^{\ddagger\sharp}$, 
        Constantino Tsallis$^*$}

\address{$^{\dagger}$Massachusetts Institute of Technology, Department of 
Nuclear Engineering, Cambridge, Massachusetts 02139 \\
$^{\ddagger}$ d'Arbeloff Laboratory for Information Systems and Technology, 
Massachusetts Institute of Technology, Department of Mechanical Engineering, 
Cambridge, Massachusetts 02139 \\
$^*$ Centro Brasileiro de Pesquisas Fisicas, Xavier Sigaud 150, 22290-180,
Rio de Janeiro-RJ, Brazil \\
$^{\sharp}$ Author to whom correspondence should be addressed }

\maketitle

\begin{abstract}

We identify a border between regular and chaotic quantum dynamics. The
border is characterized by a power law decrease in the overlap between 
a state evolved under chaotic dynamics and the same state evolved under a 
slightly perturbed dynamics. For example, the overlap decay for the quantum  
kicked top is well fitted with $[1+(q-1) (t/\tau)^2]^{1/(1-q)}$ 
(with the nonextensive entropic index $q$ 
and $\tau$ depending on perturbation strength) in the region preceding the 
emergence of quantum interference effects. This region corresponds 
to the edge of chaos for the 
classical map from which the quantum chaotic dynamics is derived.
\\
\\
PACS numbers  {03.65.Ta, 03.67.-a, 05.20.-y, 05.45.+b}
\\
\\ 
\end{abstract}

Classical chaotic dynamics is characterized by strong sensitivity 
to initial conditions. Two initially close points move apart exponentially 
rapidly as the chaotic dynamics evolve. The rate of divergence is quantified 
by the Lyapunov exponent \cite{L+L}. At the border between chaotic and 
non-chaotic regions (the `edge of chaos'), the Lyapunov exponent goes to zero.
However, it may be replaced by a generalized Lyapunov coefficient \cite{C3} 
describing power-law, rather than exponential, divergence of 
classical trajectories.

This paper identifies a characteristic signature for the edge of quantum chaos.
Quantum states maintain a constant overlap fidelity, or distance, under all 
quantum dynamics, regular and chaotic. One way to characterize quantum chaos
is to compare the evolution of an initially chosen state under the chaotic
dynamics with the same state evolved under a perturbed dynamics \cite{P1} 
\cite{P2} \cite{SC}. When the initial state is in a regular region of a 
mixed system, a system with regular and chaotic regions, the 
overlap remains close to one. When the initial state is in a chaotic zone, 
the overlap decay is exponential. This paper shows that at the
edge of quantum chaos there is a region of polynomial overlap decay. 

The Lyapunov exponent description of chaos is as follows \cite{L+L}.
If $\Delta x_0$ is the distance between two initial conditions, we define
$\xi = lim_{\Delta x_0 \rightarrow 0}(\frac{\Delta x_t}{\Delta x_0})$ to
describe how far apart two initially arbitrarily close points become at time 
$t$. Generally, $\xi(t)$ is the solution to  the differential equation 
$\frac{d\xi(t)}{dt}=\lambda_1\xi(t)$, 
such that $\xi(t) = e^{\lambda_1t}$ ($\lambda_1$ is the Lyapunov exponent). 
When the Lyapunov exponent is positive the dynamics described by $\xi(t)$ 
is strongly sensitive to initial conditions and we have chaotic dynamics.

This description of chaos works well for classical Newtonian 
mechanics but it cannot hold true for quantum 
mechanical wave functions governed by the linear Schr\"{o}dinger 
equation. Indeed, like the overlap between Liouville probability densities, 
the overlap between 
any two quantum wavefunctions is constant in time. This difficulty has led to 
the study of `quantum chaos,' the search for characteristics of quantum 
dynamics that manifest themselves as chaos in the classical realm \cite{B1}
\cite{B2} \cite{H1} \cite{BGS}. 

As a possible signature of quantum chaos, Peres \cite{P1} \cite{P2} proposed
comparing the evolution of a state under an unperturbed, $H$, and perturbed,
$H + V$, Hamiltonian for chaotic and non-chaotic dynamics. The divergence 
of the states after a time $t$ is measured via the overlap 
\begin{equation}
O(t) = |\langle\psi_u(t)|\psi_p(t)\rangle|,
\end{equation}
where $\psi_u$ is the state evolved under the unperturbed system operator and
$\psi_p$ is the state evolved under the perturbed operator.
Recent insights have sharpened the differences between chaotic and regular 
dynamics under this approach and several regimes of overlap decay behavior 
based on perturbation strength have been identified. The overlap decays for
a short time as a quadratic. After this time, for chaotic dynamics 
with weak perturbation the overlap decay is Gaussian, as expected from first 
order perturbation theory \cite{P3}\cite{Prosen}\cite{Jacq}. For stronger 
perturbations, where perturbation theory breaks down, the overlap decay 
is exponential. This occurs when the magnitude of a typical off diagonal 
element of $V$ expressed in the ordered eigenbasis of $H$ is greater than the 
average level spacing of the system, $\Delta$. The regime of exponential 
overlap decay is called the Fermi Golden Rule (FGR) regime \cite{Jacq}
\cite{Cerruti}. The rate of the exponential decay will increase with stronger 
perturbation as the perturbation strength squared until the decay rate 
reaches a value given by the classical Lyapunov exponent 
\cite{Jala}\cite{Jacq}\cite{Cas} or the bandwidth of 
$H$ \cite{Jacq}. The crossover regime from, Gaussian to exponential decay, 
has also been studied \cite{Cerruti}. We note that many of the works cited 
use $O^2$ as the fidelity. Here, we follow \cite{Prosen} and simply 
use the overlap, $O$.

For regular, non-chaotic systems the FGR regime overlap decay is a 
Gaussian, faster than the exponential decay of chaotic dynamics. This 
non-intuitive result is explained using a correlation functions formulation of 
the overlap by Prosen \cite{Prosen}. In addition, a power law decay 
$\propto t^{3/2}$ has been found for an integrable system\cite{Been}.

The initial overlap decay behavior continues until some saturation 
level \cite{Prosen}. For coherent and random pure states the 
saturation level $1/N$ for the exponential decay (in the FGR regime) 
and $2/N$ for Gaussian decay (in the weak perturbation regime), 
where $N$ is the dimension of the system Hilbert space. However, 
for eigenstates of the system and mixed random states
the saturation level increases with increasing perturbation strength 
\cite{Prosen}. 

Here we study a mixed system, a system with both chaotic and regular
dynamics. Coherent states within the regular regime are practically 
eigenstates of the system and the overlap of these states 
oscillates close to unity (see figure 1) \cite{P2}\cite{Prosen}. 
This is shown in figure 1 where the initial coherent state is centered at 
a fixed point of order one of the regular map. 
Coherent states in the chaotic regime of the system show exponential overlap
decay in the FGR regime and Gaussian overlap decay for weak perturbations. 
We show that in both the FGR and weak perturbation regimes states near the 
chaotic border have a polynomial overlap decay.

In 1988, one of us \cite{C1} introduced in statistical mechanics the 
generalized entropy
\begin{equation}
S_q = k \frac{1}{q-1}\left(1-\sum^W_{i=1}p^q_i\right)
\end{equation}
where $k$ is a positive constant, $p_i$ is the probability of finding the 
system in microscopic state, $i$, and $W$ is the number of possible 
microscopic states of the system; $q$ is the entropic index which 
characterizes the degree of the system nonextensivity. In the limit 
$q \longrightarrow 1$ we recover the usual Boltzmann entropy 
\begin{equation}
S_1 = -k\sum^W_{i=1}p_i\; lnp_i. 
\end{equation}
To demonstrate that $q$ characterizes the degree of the system nonextensivity 
it is useful to examine the $S_q$ entropy addition rule \cite{C2}. If $A$ and
$B$ are two independent systems such that the probability $p(A+B) = p(A)p(B)$
the entropy of the total system $S_q(A+B)$ is given by the following equation:
\begin{equation}
\frac{S_q(A+B)}{k}=\frac{S_q(A)}{k}+
\frac{S_q(B)}{k}+(1-q)\frac{S_q(A)S_q(B)}{k^2}.
\end{equation}
From the above equation it is realized that $q<1$ corresponds to 
superextensivity and $q>1$ to subextensivity. Using this entropy 
to generalize statistical mechanics and thermodynamics has helped explain 
many natural phenomena in a wide range of fields.  

One application of this nonextensive entropy occurs in one dimensional 
dynamical maps. As explained above, when the Lyapunov 
exponent is positive, the system dynamics is strongly sensitive to initial 
conditions and is characterized as chaotic dynamics. 
When the Lyapunov exponent is zero it has been 
conjectured \cite{C3} (and proven \cite{robledo} for the logistic map) that 
the distance between two initially arbitrarily close
points is described by $\frac{d\xi}{dt} = \lambda_{q_{sen}}\xi^{q_{sen}}$ 
leading to 
$\xi = [1+(1-q_{sen})\lambda_{q_{sen}} t]^{1/(1-q_{sen})}$ ({\it sen} stands 
for sensitivity). This requires the introduction of 
$\lambda_{q_{sen}}$ as a generalized Lyapunov coefficient. 
The Lyapunov coefficient
scales inversely with time as a power law instead of the characteristic 
exponential of a Lyapunov exponent. Thus, there exists a regime, 
$q_{sen}<1, \lambda_1 = 0, \lambda_{q_{sen}}>0$, which is weakly sensitive to 
initial 
conditions and is characterized by having power law, instead of 
exponential, mixing. This regime is called  the edge of chaos. 

The polynomial overlap decay found for initial states of a mixed system 
near the chaotic border are at the `edge of quantum chaos', 
the border between regular and chaotic quantum dynamics. This region is
the quantum parallel of the region characterized classically by 
the generalized Lyapunov coefficient.

The system studied is the quantum kicked top (QKT) \cite{H2} defined by the 
unitary operator:
\begin{equation}
U_{QKT} = e^{-i\pi J_y/2\hbar}e^{-i\alpha J_z^2/2J\hbar}.
\end{equation}
$J$ is the angular momentum of the top and $\alpha$ is the `kick' strength. 
We use a QKT with $\alpha = 3$ whose classical analog has a mixed phase space,
regions of chaotic and regular dynamics. The 
perturbed operator used is a QKT with a stronger kick strength 
$\alpha' = 3.015$. Hence, $V = e^{-i\delta J_z^2/2J\hbar}$ where 
$\delta = \alpha' - \alpha$.

\begin{figure}
\epsfig{file=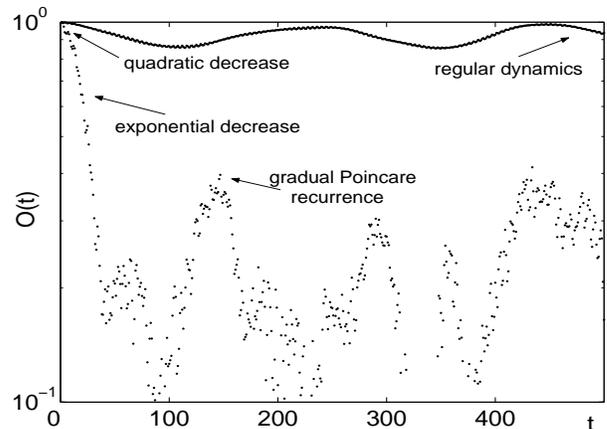, height=5.8cm, width=8cm}
\caption {Overlap versus time for initial angular momentum coherent states 
located in the chaotic region and the regular region of the quantum kicked 
top. The system has a spin 120 and is evolved under the
kicked top Hamiltonian with $\alpha = 3$ and $\alpha' = 3.015$.   
The overlap of the state in the chaotic region decreases exponentially 
with the number of iterations of the map. The state in the regular 
region is practically an eigenstate of the system and therefore 
oscillates close to unity.}
\end{figure} 

The classical kicked top is a map on the unit sphere, $x^2+y^2+z^2=1$:
\begin{eqnarray}
\begin{array}{cc}
x'= & z\\
y'= & x \;sin(\alpha z)+ y \;cos(\alpha z)\\
z'= & -x\; cos(\alpha z) + y\; sin(\alpha z).
\end{array}
\end{eqnarray}
For $\alpha = 3$ there are two fixed points of order one at the center of the 
regular regions of the map. They are located at
\begin{equation}
x_f = z_f = \pm.6294126, \;y_f = .4557187. 
\end{equation}
The regular regions of the classical phase space are seen clearly in figure 4. 

To locate the edge of quantum chaos, we work in the oo (even under a 
$180^\circ$ rotation about $x$ and odd under a $180^\circ$ about $y$; 
see page 359 of \cite{P2}) symmetry subspace of the QKT with $J = 120$ 
(in the oo subspace $N = J/2$). We set $y$ equal to $y_f$ of the 
positive fixed point and change $z$  so that the initial state can be
systematically moved closer and further from the fixed point of the map.
An initial state with a power law decrease of overlap is found at 
$z_f - .124$. The overlap decay for this state at the edge of quantum chaos 
is illustrated in figure 2 and is very well fit by the solution 
of $dO/d(t^2) = -O^{q_{rel}}/\tau_{q_{rel}}^2$  
({\it rel} stands for relaxation). Although we do 
not know how to derive this differential equation from first principles the 
numerical agreement is remarkable (see also \cite{borges}). A time-dependent
$q$-exponential expression analogous to the one shown here has recently 
been proved for the edge of chaos and other critical points of the classical 
logistic map\cite{robledo}. 
The polynomial overlap decay is the transition between the quadratic 
and exponential overlap decays. This transitory region does not appear for 
chaotic states (as shown in figure 1) and is a signature of the 
`edge of quantum chaos.'

A power law also emerges for the above initial state in the weak perturbation 
or Gaussian regime. Here we use $\alpha' = 3.0003$. The power law in this 
regime is illustrated in figure 2 and fit with the above equation. 

\begin{figure}
\epsfig{file=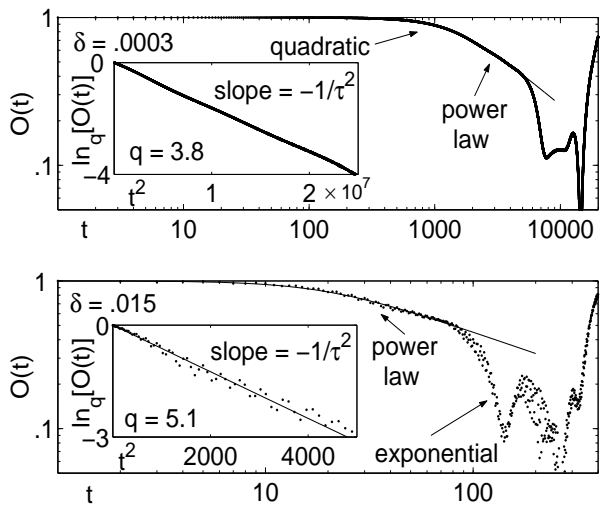, height=6.8cm, width=8cm}
\caption {Overlap versus time for an angular momentum coherent state initially
located at the border between regular and chaotic zones of the QKT of spin
120 and $\alpha = 3$. This region is called the edge of quantum chaos and
shows the expected power law decrease in overlap. The top figure is for a 
perturbation strength within the FGR regime, $\delta = .015$ and the bottom 
figure is for a perturbation strength of $\delta = .0003$, well 
below the FGR regime. On the log-log plot the power law decay region, 
from about 20-70 in the FGR regime and 2000-7000 in the Gaussian regime, 
is linear. We can fit the decrease in overlap with the expression 
$[1+(q_{rel}-1)(t/\tau_{q_{rel}})^2]^{1/(1-q_{rel})}$ where, in the FGR regime,
the entropic index $q_{rel} = 5.1$ and $\tau_{q_{rel}} = 40$ and in the 
Gaussian regime $q_{rel} = 3.8$ and $\tau_{q_{rel}} = 2500$. The insets of 
both figures show 
$\ln_{q_{rel}} O \equiv (O^{1-q_{rel}}-1)/(1-q_{rel})$ versus $t^2$; 
since $\ln_q x$ is the inverse function of 
$e_q^x \equiv [1+(1-q) \; x]^{\frac{1}{1-q}}$, this produces a straight line
with a slope $-1/\tau^2$ (also plotted).} 
\end{figure} 

The value of $q_{rel}$ remains constant at 3.8 for small perturbations until 
the critical perturbation strength, $\delta_c$, when the typical off diagonal 
elements of $V$ are larger than $\Delta$. We can approximate 
$\delta_c \simeq \sqrt{2\pi/N^3} = 5.4\times 10^{-3}$ \cite{Jacq}. 
When $\delta$ is larger than $\delta_c$, $q_{rel}$ increases. The behavior
of $q_{rel}$ and $\tau_{q_{rel}}$ versus $\delta$ can be seen in figure 3.

\begin{figure}
\epsfig{file=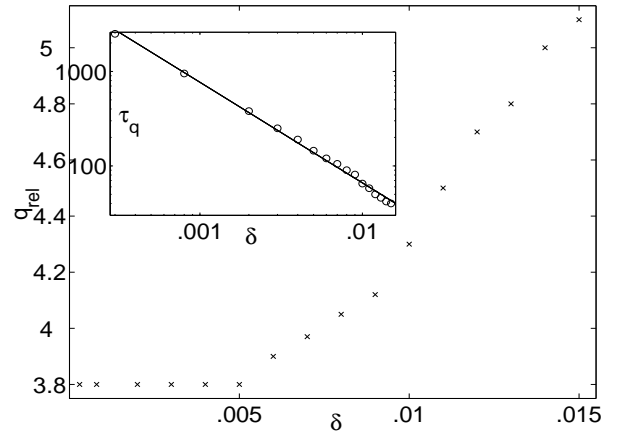, height=5.8cm, width=8cm}
\caption {$q_{rel}$ versus perturbation strength. The value of $q_{rel}$ 
remains constant at 3.8 until the critical perturbation after which it 
increases. The relationship of $\tau_{q_{rel}}$ versus perturbation strength
is shown on a log-log plot in the inset and is well fit by a line with 
slope -1.06.
} 
\end{figure} 

The location of the edge of quantum chaos for the QKT of spin 120 does not 
match up with the edge of the classical kicked top which appears at 
approximately $z_f-.2296$. This implies that classically regular regions of the
kicked top appear chaotic on the QKT. 
As $J$ is increased, the top becomes more and more classical and states 
exhibiting `edge of quantum chaos' behavior are centered closer to the 
classical value for the edge of chaos. Hence, for $J = 150, 180, 210, 240$ 
the edge is observed at $z_f - .124, .139, .151, .160$ and $.176$, 
respectively.

The edge of chaos in the quantum and classical maps are not obseved at
the same value due to the size of the angular momentum coherent state. 
The coherent state grows as $J$ decreases causing it to 
`leak out' into the chaotic region even though 
it is centered away from the chaotic border. This 
causes behavior characteristic of the edge of chaos to appear at 
different values depending on the dimension of Hilbert space. 
Figure 4 shows the wavefunctions for two values of $J$ superimposed
on the classical phase space. This gives an idea as to how large the 
wavefunction is compared to the regular region of the map.

\begin{figure}
\epsfig{file=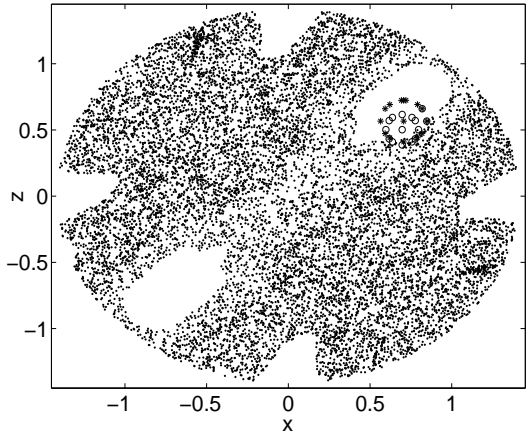, height=5.8cm, width=7cm}
\caption {Classical phase space of the kicked top with angular momentum 
coherent wavefunctions. 10000 iterations of a chaotic orbit starting 
from the point $x = .6294$, $y = .7424$, $z = .2294$. The spherical phase 
space and the ellipsoidal coherent 
states are projected onto the $x-z$ plane (only $y>0$ shown) by multiplying the
$x$ and $z$ coordinates of each point by $R/r$ where 
$R = \sqrt{2(1-|y|)}$ and $r = \sqrt{(1-y^2)}$ \protect\cite{P2}. The regular 
regions of the kicked top are clearly visible. 
Shown is a $J = 120$ wavefunction (stars) and a $J = 240$ (circles) 
wavefunction both at the edge of quantum chaos. Note that the variance of the 
$J = 120$ wavefunction is much larger than the variance of the 
$J = 240$ wavefunction. Hence, the behavior characteristic of the 
edge of quantum chaos appears further from the fixed point of the 
classical map.}
\end{figure} 

In the region of $J$ values that we have explored, no significant changes 
have been detected for $q_{rel}$, because the $\delta_c$ changes only  
slightly. However, $\tau_{q_{rel}}$ decreases with increasing $J$.

To conclude, we have located a region on the border of chaotic and 
non-chaotic quantum dynamics. Quantum states located in this 
region exhibit a power-law decrease in overlap as opposed to the exponential
overlap decay exhibited by fully chaotic quantum dynamics. The classical 
parallel to this region is the border between 
regular and chaotic classical dynamics which is characterized by the 
generalized Lyapunov coefficient.

The authors would like to acknowledge C. Anteneodo and J. Emerson for 
useful remarks. This work was supported by DARPA/MTO through ARO grant 
DAAG55-97-1-0342.

\end{document}